\documentstyle[12pt,a4,graphicx]{article}
\begin{document}
\,\,\,\,

\centerline{\bf NOVEL EFFECTS IN HIGH-T$_{\bf C}$ GRANULAR
SUPERCONDUCTORS} \centerline{\bf PREDICTED BY A MODEL OF 3D JOSEPHSON
JUNCTION ARRAYS} \vspace{15mm}

\centerline{\bf SERGEI SERGEENKOV} \vspace{8mm}

\centerline{Bogoliubov Laboratory of Theoretical Physics,}
\centerline{Joint Institute for Nuclear Research, Dubna, Russia}
\vspace{20mm}

\leftline{\bf 1. INTRODUCTION} \vspace{5mm}

Despite the fact that Josephson Junction Arrays (JAA) have been
actively studied for decades, they continue to contribute to the
variety of intriguing and peculiar phenomena (both fundamental and
important for potential applications) providing at the same time a
useful tool for testing new theoretical ideas. To give just a few
recent examples, it is sufficient to mention paramagnetic Meissner
effect (PME)[1-5] as well as the recently introduced thermophase
[6,7] and piezophase [8] effects suggesting, respectively, a direct
influence of a thermal gradient and an applied stress on phase
difference between the adjacent grains. At the same time, an
artificially prepared islands of superconducting grains [4,5,9,10],
well-described by the various models of JJAs, proved useful in
studying the charging effects in these systems, ranging from Coulomb
blockade of Cooper pair tunneling and Bloch oscillations [11,12] to
propagation of quantum ballistic vortices [13].

The present paper reviews some of the recently suggested novel
effects [14-16] which should manifest themselves either in
weak-links-bearing superconductors or in an artificially prepared
JJAs. In Section 2 we consider the appearance of an electric-field
induced magnetization (analog of the so-called magnetoelectric
effect) in a model granular superconductor. The dual effect, that is
an appearance of a magnetic field induced electric polarization is
studied in Section 3. Finally, in Section 4 we discuss a possibility
of two other interesting effects which are expected to occur in a
granular material under mechanical loading. Specifically, we predict
the existence of stress induced paramagnetic moment in zero applied
magnetic field (Josephson piezomagnetism) and study its influence on
a low-field magnetization (leading to a mechanically induced PME).

\vspace{8mm} \leftline{\bf 2. MAGNETOELECTRIC EFFECT} \vspace{5mm}

To account for the unusual behavior of the critical current under the
influence of a strong enough applied electric field observed in
high-$T_c$ superconductos (HTCS) [17-20], a possibility of the
superconducting analog of the so-called magnetoelectric effect (MEE)
in JJAs has been recently suggested [15,21]. This effect is similar
to (but physically different from) the MEE seen in magnetic
ferroelectrics (like antiferromagnetic $BiFeO_3$) [22]. It has also
been discussed in the context of some exotic nonmagnetic normal metal
conductors, with the symmetry of a mirror isomer [23], and
pyroelectric superconductors (where the supercurrent passing through
a metal of a polar symmetry is assumed to be accompanied by spin
polarization of the carriers [24]). The effect discussed here entails
the electric field generation of a magnetic moment (paramagnetic in
small to diamagnetic for large electric fields) due to
superconducting currents that circulate between grains. The basic
physical reason for the appearance of this effect is that the applied
electric field induces a magnetization $M$, that changes its sign
($M\to -M$) under a time-parity transformation. In the case of
weakly-coupled superconducting grains, the phenomenological reason
for the time-parity violation comes from the fact that the total free
energy $F$ for the material exhibiting both magnetic and electric
properties contains the term $\alpha_{mn}f_{mn}(\vec E,\vec H)$, with
the coefficients $\alpha_{mn}\neq 0$ (here $\{m,n\}=x,y,z$). In the
standard MEE, the function $f_{mn}\equiv E_mH_n$, which leads to the
corresponding {\it linear} effect, which is either an electric-field
induced magnetization $M_m(\vec E)\equiv \partial F/\partial
H_m=\alpha _{mn}E_n$ (in zero magnetic field) or a magnetic-field
induced polarization $P_m(\vec H)\equiv \partial F/\partial
E_m=\alpha _{mn}H_n$ (in zero electric field). Since $H$ (and $M$)
changes its sign under time-parity transformation while $E$ (and $P$)
remains unchanged, an electric field induced magnetization will break
time-parity symmetry even in zero magnetic field applied. As we show
below, in our case the symmetry breaking term in $F$, represented by
a nonzero coefficient $\alpha_{mn}$, has a more general {\it
nonlinear} form for $f_{mn}$. We recall that the standard linear MEE
can appear when an external electric field $\vec E$ interacts with an
inner magnetic field $\vec h_{DM}$ of the Dzyaloshinskii-Moriya (DM)
type [25]. The DM interaction leads to a term, apart from the
standard isotropic term
 in the Heisenberg Hamiltonian, of the form
$H_{DM}= \sum _{i,j} \vec D_{i,j}\cdot (\vec S_i\wedge \vec S_j)$,
where $\vec S_i$ is a Heisenberg spin and the constant vector $\vec
D_{i,j}$ arises from the spin-orbit coupling.
 An analogous
situation occurs in our case, as we describe below. To see how we can
get a nonzero $\alpha$, or equivalently a DM type interaction in a
granular superconductor, we model a HTCS ceramic sample by a {\it
random} three-dimensional (3D) overdamped Josephson junction array.
This model has proven to be useful in describing the metastable
magnetic properties of HTCS [26,27]. In thermodynamic equilibrium,
this model has a Boltzmann factor with a random 3D-XY model
Hamiltonian. Specifically, the general form of the Hamiltonian
(describing both DC and AC effects) reads
\begin{equation}
{\cal H}(t)=\sum_{i,j}J(r_{i,j})[1- \cos \phi_{i,j}(t)]
\end{equation}
Here $\{i\}=\vec {r}_i$ is a 3D lattice vector; $J(r_{i,j})$ is the
Josephson coupling energy, with $\vec r_{i,j}=\vec r_i-\vec r_j$ the
separation between the grains; the  gauge invariant phase difference
is defined as
\begin{equation}
\phi _{i,j}(t)=\phi _{i,j}(0)-A_{i,j}(t),
\end{equation}
where $\phi _{i,j}(0)=\phi _i-\phi _j$ with $\phi_i$ being the phase
of the superconducting order parameter; $A_{i,j}(t)$ is
(time-dependent, in general) frustration parameter, defined as
\begin{equation}
A_{i,j}(t)=\frac{2\pi}{\Phi_ 0}\int_i^j\vec A(\vec r,t)\cdot d{\vec
l},
\end{equation}
with $\vec A(\vec r,t)$ the (space-time dependent) electromagnetic
vector potential which involves both external fields and the electric
and magnetic possible  self-field effects (see below); $\Phi_ 0=h/2e$
is the quantum of flux, with $h$ Planck's constant, and $e$ the
electronic charge. Expanding the cosine term, and using trigonometric
identities we can explicitly rewrite the above Hamiltonian as ${\cal
H}=-\sum _{i,j}J[\cos (A_{i,j})\vec S_i\cdot\vec S_j-\sin (A_{i,j})
{\hat k}\cdot {\vec S_i}\wedge {\vec S_j}]$, where the two-component
XY spin vector is defined as $\vec S_i\equiv (\cos\phi_i,
\sin\phi_i)$, and $\hat k$ is a unit vector along the z-axis [28]. We
see that the second term in this Hamiltonian has the same form as in
the DM contribution, and thus we can surmise that the time parity
will be broken by applying an external field (electric or magnetic)
to the granular system. To bring to the fore this possibility, we
show below in a simple but yet nontrivial model that this is indeed
the case.

There are different types of $A_{i,j}$ randomness that can be
considered [27]. For simplicity, in the present paper, we consider a
long-range interaction between grains (assuming $J(r_{i,j})=J$) and
model the true short-range behavior of a ceramics sample through the
randomness in the position of the superconducting grains in the array
(using the exponential distribution law $f_r(r_{i,j})$, see below).
Here we restrict our consideration to the case of an external
electric field only but it can be shown that the scenario suggested
will also carry through when applying an external magnetic field
(which will induce another time-parity breaking phenomenon in the
granular material; namely, magnetic field induced electric
polarizability, see the next Section). Besides, in what follows we
also ignore the role of Coulomb interaction effects assuming that the
grain's charging energy $E_C\ll J$ (where $E_C=e^2/2C$, with C the
capacitance of the junction).

In the case of a granular material, we show here that the
corresponding time-parity breaking DM internal field can be related
to the electric field induced magnetic moment produced by the
circulating Josephson currents between the grains. As is known
[29,30], a constant electric field $\vec E$ applied to a single
Josephson junction (JJ) causes a time evolution of the phase
difference. In this particular case Eq.(2) reads
\begin{equation}
\phi _{i,j}(t)=\phi _{i,j}(0)+\frac{2e}{\hbar}\vec E\cdot \vec
r_{i,j}t
\end{equation}
The resulting  AC superconducting current in the junction is
\begin{equation}
I_{i,j}^s(t)=\frac{2eJ}{h} \sin \phi _{i,j}(t)
\end{equation}
If, in addition to the external electric field $\vec E$, the network
of superconducting grains is under the influence of an applied
magnetic field $\vec H$, the frustration parameter $A_{i,j}(t)$ in
Eq.(3) takes the following form
\begin{equation}
A_{i,j}(t)=\frac{\pi}{\Phi _0}(\vec H\wedge \vec R_{i,j})\cdot \vec
r_{i,j}- \frac{2\pi}{\Phi _0}\vec E\cdot \vec r_{i,j}t
\end{equation}
Here, $\vec R_{i,j}=(\vec r_i+\vec r_j)/2$, and we have used the
conventional relationship between the vector potential $\vec A$ and a
constant magnetic field $\vec H=rot \vec A$ (with $\partial \vec
H/\partial t=0$), as well as a homogeneous electric field $\vec
E=-\partial \vec A/\partial t$ (with $rot \vec E=0$). In the type II
HTCS the magnetic self-field effects for the array as a whole are
expected to be negligible [31]. The grains themselves are in fact
larger than the London penetration depth and we must then have that
the corresponding Josephson penetration length must be much larger
than the grain size (since the self-induced magnetic fields can in
principle be quite pronounced for large-size junctions even in zero
applied magnetic fields [14]). Specifically, this  is justified for
short junctions with the size $d\ll \lambda _J$, where $\lambda
_J=\sqrt{\Phi _0/4\pi \mu _0j_c \lambda _L}$ is the Josephson
penetration length  with $\lambda _L$ being the grain London
penetration depth and $j_c$ its Josephson critical current density.
In particular, since in HTCS $\lambda _L\simeq 150nm$, the above
condition will be fulfilled for $d\simeq 1\mu m$ and $j_c\simeq
10^{4}A/m^2$ which are the typical parameters for HTCS ceramics [29].
Likewise, to ensure the uniformity of the applied electric field, we
also assume that $d\ll \lambda _E$, where $\lambda _E$ is an
effective electric field penetration depth [20,21].

When the AC supercurrent $I_{i,j}^s(t)$ (defined by Eqs.(2), (5) and
(6)) circulates around a set of grains, that form a random area
plaquette, it induces a random AC magnetic moment $\vec \mu _s(t)$ of
the Josephson network [26]
\begin{equation}
\vec \mu _s(t)\equiv \left [\frac{\partial {\cal H}}{\partial \vec H}
\right ]_{\vec H=0}= \sum_{i,j}I_{i,j}^s(t)(\vec r_{i,j}\wedge \vec
R_{i,j})
\end{equation}
Notice that in the MEE-like effect discussed here for a granular
superconductor, the electric-field induced magnetic moment in the
system is still present in zero applied magnetic field due to the
phase coherent currents between the weakly-coupled superconducting
grains.

To consider the essence of the superconducting analog of MEE, we
assume that in a {\it zero electric field} the phase difference
between the adjacent grains $\phi _{i,j}(0)=0$ which corresponds to a
fully coherent state of the array. In this particular case, the
electric-field induced averaged magnetization reads
 \begin{equation}
\vec M_s(\vec E)\equiv \overline{\vec \mu_s (t)}=
\frac{1}{\tau}\int\limits_{0}^
 {\tau }dt \int\limits_{0}^{\infty }d\vec r_{i,j}d\vec R_{i,j}
f(\vec r_{i,j}, \vec R_{i,j}) \vec \mu _s(t),
 \end{equation}
where $\tau$ is the electronic relaxation scattering time, and $f$ is
the joint probability distribution function (see below).

To obtain an explicit expression for the electric-field dependent
magnetization, we consider a site positional disorder that allows for
small random radial displacements. Namely, the sites in a 3D cubic
lattice are assumed to move from their equilibrium positions
according to the normalized (separable) distribution function
\begin{equation}
 f(\vec r_{i,j}\vec R_{i,j})\equiv f_{r}(\vec r_{i,j})f_{R}(\vec R_{i,j})
\end{equation}
It can be shown that the main qualitative results presented here do
not depend on the particular choice of the probability distribution
function. For simplicity here we assume an exponential distribution
law for the distance between grains, $f_r(\vec r)=f(x_1)f(x_2)f(x_3)$
with $f_r(x_j)=(1/d)e^{-x_j/d}$, and a short range distribution for
the dependence of the center-of-mass probability $f_R(\vec R)$
(around some constant value $D$). The specific form of the latter
distribution will not affect the qualitative nature of the final
result. (Notice that in fact the former distribution function
$f_r(\vec r)$ reflects a short-range character of the Josephson
coupling in granular superconductor. Indeed, according to the
conventional picture [32] the Josephson coupling $J(\vec r_{ij})$ can
be  assumed to vary exponentially with the distance $\vec r_{ij}$
between neighboring grains, i.e., $J(\vec r_{ij})=Je^{-\vec \kappa
\cdot \vec r_{ij}}$. For isotropic arrangement of identical grains,
with spacing $d$ between the centers of adjacent grains, we have
$\vec \kappa =(\frac{1}{d},\frac{1}{d},\frac{1}{d})$ and thus $d$ is
of the order of an average grain size.) Taking the applied electric
field along the $x$-axis, $\vec E=(E_x,0,0)$, we get finally
\begin{equation}
 M_z(E_x)=\frac{B_z(E_x)}{\mu _0}-H_z(E_x),
 \end{equation}
for the induced transverse magnetization (along the $x_3=z$-axis),
where
 \begin{equation}
 B_z(E_x)=\mu _0M_0\frac{E_x/E_0}{1+(E_x/E_0)^2},
 \end{equation}
 and
 \begin{equation}
 H_z(E_x)=M_0\left (\frac{E_0}{E_x}\right ) \log \sqrt{1+
\left (\frac{E_x}{E_0}\right )^2},
 \end{equation}
stand for the electric-field induced  magnetic induction $B_z(E_x)$
and magnetic field $H_z(E_x)$, respectively. The induced Josephson
current $I(E_x)$ is simply given by Ampere's law $I(E_x)=H_z(E_x)d$.
In these equations, $M_0=2\pi eJNdD/\hbar $, with $N$ the total
number of grains and $E_0=\hbar /2de\tau$. Eq.(10) is the main result
of this Section, which we proceed to analyze below. As is seen from
Eq.(10), the behavior of the magnetization in the applied electric
field is determined by the competition between the two contributions,
the magnetic induction $B_z(E_x)$ and the current induced magnetic
field $H_z(E_x)$ (or the corresponding Josephson current $I(E_x)$).
Namely, below a critical (threshold) field $E_c\approx 1.94\, E_0$
where $B_z(E_x)>\mu _0H_z(E_x)$, a paramagnetic phase of the MEE is
due to the modification of the magnetic induction in the applied
electric field. On the other hand above the threshold field (when
$E_x$ becomes larger than $E_c$) the Josephson current $I(E_x)$
induced contribution starts to prevail, leading to the appearance of
the diamagnetic signal (seen as a small negative part  of the induced
magnetization in Fig.1).
\begin{figure*}[t]
\centerline{\includegraphics[width=12.cm,clip=true]{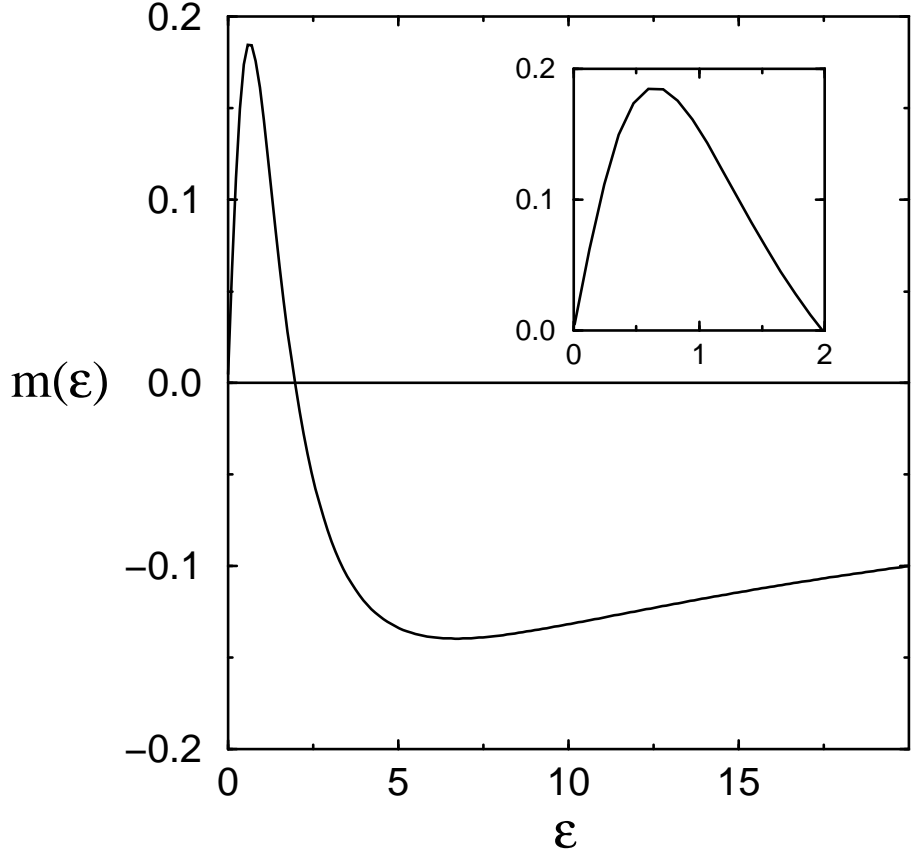}}
\caption{The induced magnetization $m(\epsilon)=M_z/M_0$ as a
function of the normalized applied electric field $\epsilon
=E_x/E_0$, according to Eq.(10). The inset shows a blow up of the
paramagnetic region of $m(\epsilon)$ (exhibiting its linear
superconducting magnetoelectric effect regime).}
\end{figure*}
 Such an electric field induced
paramagnetic-to-diamagnetic transition has been actually observed
[17-19] for behavior of the critical current in ceramic HTCS and it
was attributed [20] to a proximity-mediated enhancement of the
superconductivity in a granular material in a strong enough electric
field. Explicitly, the critical current in a $YBCO$ sample was found
to reach a maximum  at $E=4\times 10^7 V/m$. To relate this
experimental value with the model parameters, first of all, we need
to estimate an order of magnitude of the relaxation time $\tau$ in a
zero applied electric field. This will provide an {\it upper} limit
for the relevant $\tau$-distribution in our system. It is reasonable
to connect zero-field $\tau \equiv \tau (0)$ with the Josephson
tunneling time [29] $\tau _J =(R_0/R_n)(\hbar /J)$ (where
$R_0=h/4e^2$, and $R_n$ is the normal state resistance between
grains). Typically, for HTCS ceramics $J/k_B\simeq 90K$ and
$R_n/R_0\simeq 10^{-3}$, so that $\tau _J\simeq 10^{-10}s$. At the
same time, at high enough electric fields where the MEE becomes
strongly nonlinear, we can expect quite a tangible decrease of the
relaxation time. Indeed, for an average grain size $d\approx 1\mu m$,
the characteristic field $E_0=4\times 10^7 V/m$ (which corresponds to
the region where a prominent enhancement of the critical current was
observed [17-19]) introduces a substantially shorter relaxation time
$\tau (E_0)=\hbar /2deE_0\simeq 10^{-16}s$, in agreement with
observations.

To estimate the relative magnitude of the superconducting analog of
the MEE predicted here, we can compare it with the normal (Ohmic)
contribution to the magnetization
\begin{equation}
\vec M_n(\vec E)\equiv \overline{\vec \mu_n
(t)}=\frac{1}{\tau}\int\limits_{0}^ {\tau }dt \int\limits_{0}^{\infty
}d\vec r_{i,j}d\vec R_{i,j}f(\vec r_{i,j}, \vec R_{i,j}) \vec \mu _n,
\end{equation}
where
\begin{equation}
\vec \mu _n=\sum_{i,j}I_{i,j}^n(\vec r_{i,j}\wedge  \vec R_{i,j})
\end{equation}
Here $I_{i,j}^n=V_{i,j}/R_n$ is the normal current component  due to
the applied electric field $\vec E$, with $V_{i,j}=\vec E\cdot \vec
r_{i,j}$ being the induced voltage, and $R_n$ the normal state
resistance between grains. As a result, the normal state contribution
(for $\vec E$ along the $x$-axis) reads $M_n=\alpha _nE_x$, with
$\alpha _n=\pi d^2DN/R_n$. Similarly, according to Eq.(10), the low
field contribution to the superconducting MEE gives $M_s\simeq \alpha
_sE_x$ with $\alpha _s=2\pi e^2JN\tau (0)d^2D/\hbar ^2$. Thus, at low
enough applied fields (when $E_x\ll E_0$)
\begin{equation}
\frac{\alpha _s}{\alpha _n}\simeq \frac{\tau (0)}{\tau _J},
\end{equation}
where $\tau _J =(R_0/R_n)(\hbar /J)$ with $R_0=h/4e^2$. According to
our previous discussion on the relevant relaxation-time distribution
spectrum in our model system, we may conclude that $\tau (0)\leq \tau
_J$. So, we arrive at the following ratio between the coefficients of
the superconducting to normal MEEs, namely $\alpha _s/\alpha _n\leq
1$. The above estimate of the weak-links induced MEE (along with its
rather specific field dependence, see Fig.1) suggests quite an
optimistic possibility to observe the predicted effect experimentally
in HTCS ceramics or in a specially prepared system of arrays of
superconducting grains.

We note that in the present analysis we have not explicitly
considered the polarization effects (due to the interaction between
the applied electric field and the grain's charges) which may become
important at high enough fields (or for small enough grains), leading
to more subtle phenomena (like Coulomb blockade and reentrant-like
behavior) that will demand the inclusion of charging energy effects
in the analysis.

\vspace{8mm} \leftline{\bf 3. MAGNETIC FIELD INDUCED CHARGING
EFFECTS} \vspace{5mm}

This Section addresses a related phenomenon which is actually dual to
the above-discussed analog of magnetoelectric effect. Specifically,
we analyze a possible appearance of a non-zero electric polarization
and the related change of the charge balance in the system of
weakly-coupled superconducting junctions (modelled by the random $3D$
JJAs) under the influence of an applied magnetic field. Since the
field-induced effects considered in this Section are expected to
manifest themselves in high enough applied magnetic fields (with a
nearly homogeneous distribution of magnetic flux along the junctions)
and as long as the Josephson penetration length $\lambda _J$ exceeds
the characteristic size of the Josephson network $d$ (which is
related to the projected junction area $S$, where the field
penetration actually occurs, as follows $S=\pi d^2$), the Josephson
current-induced "self-field" effects (which are important for a
large-size junctions and/or small applied magnetic fields) may be
safely neglected [33]. Besides, it is known [34] that in discrete
JJAs pinning (by a single junction) actually concurs with the
"self-field" effects. Specifically, it was found [34] that the ratio
$d/\lambda _J$ is related to the dimensionless pinning strength
parameter $\beta$ as $d/\lambda _J=\sqrt{\beta}$ suggesting that a
weak pinning regime (with $\beta \ll 1$) simultaneously implies a
smallness of the "self-field" effect and vice versa. And since
artificially prepared Josephson networks allow for more flexibility
in varying the experimentally-controlled parameters, it is always
possible to keep the both above effects down by appropriately tuning
the ratio $d/\lambda _J$. Typically, in this kind of experiments
$S=0.01-0.1\mu m^2$ and $d\ll \lambda _J$.

In what follows, we are interested in the magnetic field induced
behavior of the electric polarization in a $3D$ JJA at zero
temperature. Recall that a conventional (zero-field) pair
polarization operator within the model under discussion reads [30]
\begin{equation}
\vec p=\sum_{i=1}^Nq_i \vec r_i,
\end{equation}
where $q_i =-2en_i$ with $n_i$ the pair number operator, and $r_i$ is
the coordinate of the center of the grain.

In view of Eqs.(1)-(6), and taking into account a usual
"phase-number" commutation relation, $[\phi _i,n_j]=i\delta _{i,j}$,
the evolution of the pair polarization operator is determined via the
equation of motion
\begin{equation}
\frac{d\vec p}{dt}=\frac{1}{i\hbar}\left[ \vec p,{\cal H}\right ]
=\frac{2e}{\hbar }\sum_{ij}^NJ\sin \phi _{i,j}(\vec H)\vec r_{i,j}
\end{equation}
Resolving the above equation, we arrive at the following net value of
the magnetic-field induced polarization (per grain)
\begin{equation}
\vec P(\vec H)\equiv \frac{1}{N}\overline {<\vec p(t)>}=\frac{2eJ}
{\hbar \tau N} \int\limits_{0}^ {\tau }dt \int
\limits_{0}^{t}dt'\sum_{ij}^N <\sin \phi _{i,j}(\vec H)\vec r_{i,j}>,
\end{equation}
where $<...>$ denotes a configurational averaging over the grain
positions, while the bar means a temporal averaging (with a
characteristic time $\tau$). To consider a field-induced polarization
only, we assume that in a zero magnetic field, $\vec P\equiv 0$,
implying $\phi _{i,j}(0)\equiv0$.

Since, as usual, the main qualitative results presented here do not
depend on the particular choice of the probability distribution
function, for a change, the following law $f(\vec r, \vec R)=f_r(\vec
r)f_R(\vec R)$ with $f_r(\vec r)=\frac{1}{(2\pi
d^2)^{3/2}}e^{-(x^2+y^2+z^2)/2d^2}$ and $f_R(\vec R)=\delta
(X-d)\delta (Y-d)\delta (Z-d)$ will be assumed. As a result, we
observe that the magnetic field $\vec H=(0,0,H_z)$ (applied along the
$z$-axis) will induce a non-vanishing longitudinal (along $x$-axis)
electric polarization
\begin{equation}
P_x(H_z)=P_0G(H_z/H_0),
\end{equation}
with
\begin{equation}
G(z)=ze^{-z^2}
\end{equation}
Here $P_0=ed\tau J/\hbar $, $H_0=\Phi _0/S$ with $S=\pi d^2$ being an
average projected area of a single junction, and $z=H_z/H_0$.
\begin{figure*}[t]
\centerline{\includegraphics[width=9.cm,clip=true]{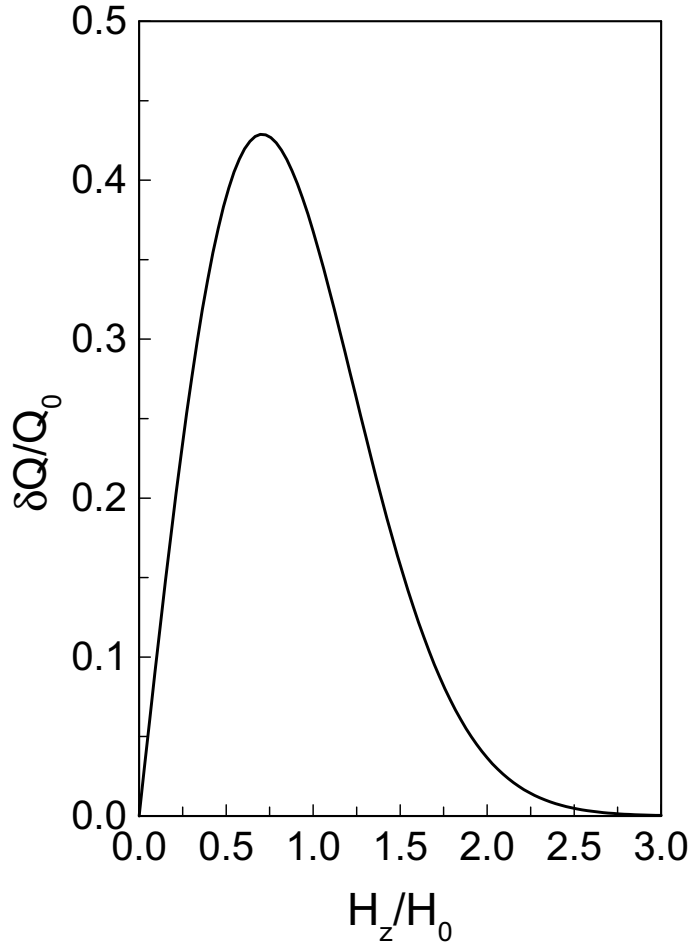}}
\caption{The behavior of the induced effective charge $\delta Q/Q_0$
 in applied magnetic field $H_z/H_0$, according to Eq.(21).}
\end{figure*}
For small applied fields ($z\ll 1$), the induced polarization
$P_x(H_z)\approx \alpha _0H_z$, where $\alpha _0=e\tau dJ/\hbar H_0$
is the so-called linear magnetoelectric coefficient [35]. However, as
we mentioned in the very beginning, to correctly describe any induced
effects in very small external fields, both Josephson junction
pinning and "self-field" effects have to be taken into account. At
the same time, in view of Eq.(16) the induced polarization is related
to the corresponding change of the effective charge $\delta Q\equiv
(1/d) \sum_i<q_ix_i>$ in applied magnetic field as follows
\begin{equation}
\delta Q(H_z)=\frac{P_x(H_z)}{d}=Q_0G(H_z/H_0),
\end{equation}
where $Q_0=e\tau J/\hbar$.

It is of interest also to consider the related field behavior of the
effective flux capacitance $\delta C\equiv \tau d\delta Q(H_z)/d\Phi
$ which in view of Eq.(21) reads
\begin{equation}
\delta C(H_z)=C_0\left( 1-2\frac{H_z^2}{H_0^2}\right
)e^{-H_z^2/H_0^2},
\end{equation}
where $\Phi =SH_z$, and $C_0=\tau Q_0/\Phi _0$. Figure 2 shows the
behavior of the induced effective charge $\delta Q/Q_0$ in applied
magnetic field $H_z/H_0$. As is seen, at $H_z/H_0\approx 0.8$ the
effective charge reaches its maximum (while the capacitance changes
its sign at this field), suggesting a significant redistribution of
the junction charge balance in a model system under the influence of
an applied magnetic field, near the Josephson critical field $H_0$.
Note that a somewhat similar behavior of the magnetic field induced
charge (and related capacitance) has been observed in 2D electron
systems [36].

Taking $\tau =10^{-10}s$ for the Josephson relaxation time (which is
related to the Josephson plasma frequency, $\omega _p\simeq \tau
^{-1}$, known to be the characteristic frequency of the system for
$E_C\ll J$ regime, with $E_C$ being the Coulomb grain's charge
energy; typically $\omega _p=10^9-10^{11}Hz$) and $J/k_B=90K$ for a
zero-temperature Josephson energy in $YBCO$ ceramics, we arrive at
the following estimates of the effective charge $Q_0\approx
10^{-16}C$, flux capacitance $C_0\approx 10^{-11}F$, the equivalent
current $I_0\sim Q_0/\tau \approx 10^{-6}A$, and voltage $V_0\sim
Q_0/C_0\approx 10^{-5}V$. We note that the above set of estimates
fall into the range of parameters used in typical experiments to
study the charging effects both in single JJs (with the working
frequency from RF range of $\omega \simeq 10 GHz$ used to stimulate
the system [13]) and JJAs [37] suggesting thus quite an optimistic
possibility to observe the above-predicted field induced effects
experimentally, using a specially prepared system of arrays of
superconducting grains.

\vspace{8mm} \leftline{\bf 4. STRESS INDUCED PARAMAGNETIC MEISSNER
EFFECT} \vspace{5mm}

The possibility to observe tangible piezoeffects in mechanically
loaded grain boundary Josephson junctions (GBJJs) is based on the
fact that under plastic deformation, grain boundaries (GBs) (which
are the natural sources of weak links in HTCS), move rather rapidly
via the movement of the grain boundary dislocations (GBDs) comprising
these GBs [38-40]. Using the above evidence, in Ref.8 a {\it
piezophase} response of a single GBJJ (created by GBDs strain field
$\epsilon _d$ acting as an insulating barrier of thickness $l$ and
height $U$ in a $SIS$-type junction with the Josephson energy
$J\propto e^{-l\sqrt{U}}$) to an applied stress was considered. To
understand how piezoeffects manifest themselves through GBJJs, let us
invoke an analogy with the so-called {\it thermophase effect} [6,7]
(a quantum mechanical alternative for the conventional thermoelectric
effect) in JJs. In essence, the thermophase effect assumes a direct
coupling between an applied temperature drop $\Delta T$ and the
resulting phase difference $\Delta \phi$ through a JJ. When a rather
small temperature gradient is applied to a JJ, an entropy-carrying
normal current $I_n=L_n\Delta T$ (where $L_n$ is the thermoelectric
coefficient) is generated through such a junction. To satisfy the
constraint dictated by the Meissner effect, the resulting
supercurrent $I_s=I_c\sin [\Delta \phi ]$ (with $I_c=2eJ/h$ being the
Josephson critical current) develops a phase difference through a
weak link. The normal current is locally canceled by a counterflow of
supercurrent, so that the total current through the junction
$I=I_n+I_s=0$. As a result, supercurrent $I_c\sin [\Delta \phi
]=-I_n=-L_n\Delta T$ generates a nonzero phase difference leading to
the linear thermophase effect [6,7] $\Delta \phi \simeq -L_{tp}\Delta
T$ with $L_{tp}=L_n/I_c(T)$.

By analogy, we can introduce a {\it piezophase effect} (as a quantum
alternative for the conventional piezoelectric effect) through a JJ
[8,16]. Indeed, a linear conventional piezoelectric effect relates
induced polarization $P_n$ to an applied strain $\epsilon$ as [35]
$P_n=d_n\epsilon$, where $d_n$ is the piezoelectric coefficient. The
corresponding normal piezocurrent density is $j_n=dP_n/dt=d_n\dot
{\epsilon}$ where $\dot {\epsilon}(\sigma )$ is a rate of plastic
deformation which depends on the number of GBDs of density $\rho$ and
a mean dislocation rate $v_d$ as follows [41] $\dot {\epsilon}(\sigma
)=b\rho v_d(\sigma )$ (where $b$ is the absolute value of the
appropriate Burgers vector). In turn, $v_d(\sigma )\simeq v_0(\sigma
/\sigma _m)$. To meet the requirements imposed by the Meissner
effect, in response to the induced normal piezocurrent, the
corresponding Josephson supercurrent of density $j_s=dP_s/dt=j_c\sin
[\Delta \phi ]$ should emerge within the contact. Here $P_s=-2enb$ is
the Cooper pair's induced polarization with $n$ the pair number
density, and $j_c=2ebJ/\hbar V$ is the critical current density. The
neutrality conditions ($j_n+j_s=0$ and $P_n+P_s=const$) will lead
then to the linear piezophase effect $\Delta \phi \simeq -d_{pp}\dot
{\epsilon}(\sigma )$ (with $d_{pp}=d_n/j_c$ being the piezophase
coefficient) and the concomitant change of the pair number density
under an applied strain, viz., $\Delta n(\epsilon )=d_{pn}\epsilon$
with $d_{pn}=d_n/2eb$.

To adequately describe magnetic properties of a granular
superconductor, we again employ a model of {\it random}
three-dimensional (3D) overdamped Josephson junction array which is
based on the familiar tunneling Hamiltonian given by Eq.(1) (see
Section 2). According to the above-discussed scenario, under
mechanical loading the superconducting phase difference will acquire
an additional contribution $\delta \phi _{i,j}(\sigma ) = -B\vec
\sigma \cdot \vec r_{i,j}$, where $B=d_n\dot{\epsilon} _0/\sigma
_mj_cb$ with $\dot{\epsilon} _0=b\rho v_0$ being the maximum
deformation rate and the other parameters defined earlier. If, in
addition to the external loading, the network of superconducting
grains is under the influence of an applied frustrating magnetic
field $\vec H$, the total phase difference through the contact reads
\begin{equation}
\phi _{i,j}(\vec H, \vec \sigma )=\phi ^0_{i,j}+\frac{\pi}{\Phi _0}
(\vec r_{i,j}\wedge \vec R_{i,j})\cdot \vec H-B\vec \sigma \cdot \vec
r_{i,j}
\end{equation}
Once again, to neglect the influence of the self-field effects in a
real material, the corresponding Josephson penetration length
$\lambda _J$ must be much larger than the junction (or grain) size.
Likewise, to ensure the uniformity of the applied stress $\sigma$, we
also assume that $d\ll \lambda _{\sigma}$, where $\lambda _{\sigma}$
is a characteristic length over which $\sigma$ is kept homogeneous.

When the Josephson supercurrent $I_{i,j}^s=I_c\sin \phi _{i,j}$
circulates around a set of grains, it induces a random magnetic
moment $\vec \mu _s$ of the Josephson network which results in the
stress induced net magnetization (Cf. Eq.(8))
\begin{equation}
\vec M_s(\vec H,\vec \sigma)\equiv <\vec \mu_s>=
\int\limits_{0}^{\infty }d\vec r_{i,j}d\vec R_{i,j} f(\vec r_{i,j},
\vec R_{i,j}) \vec \mu _s
\end{equation}
To capture the very essence of the superconducting piezomagnetic
effect, in what follows we assume for simplicity that an {\it
unloaded sample} does not possess any spontaneous magnetization at
zero magnetic field (that is $M_s(0,0)=0$) and that its Meissner
response to a small applied field $H$ is purely diamagnetic (that is
$M_s(H,0)\simeq -H$). According to Eq.(23), this condition implies
$\phi ^0_{i,j}=2\pi m$ for the initial phase difference with $m=0,\pm
1, \pm 2,..$.

Taking the applied stress along the $x$-axis, $\vec
\sigma=(\sigma,0,0)$, normally to the applied magnetic field $\vec
H=(0,0,H)$, and assuming an exponential distribution law for the
distance between grains, $f_r(\vec r)=f(x)f(y)f(z)$ with
$f(x_j)=(1/d)e^{-x_j/d}$ (see Section 2 for detailes), we get finally
\begin{equation}
M_s(H,\sigma )=-M_0(\sigma )\frac{H_{tot}(H,\sigma )/H_0}
{[1+H^2_{tot}(H,\sigma )/H^2_0]^2},
\end{equation}
for the induced transverse magnetization, where $H_{tot}(H,\sigma
)=H-H^{*}(\sigma )$ is the total magnetic field with $H^{*}(\sigma
)=[\sigma /\sigma _0(\sigma )]H_0$ being a stress-induced
contribution. Here, $M_0(\sigma )=I_c(\sigma )SN/V$ with $S=\pi dD$
being a projected area around the JJ, $H_0=\Phi _0/S$, and $\sigma
_0(\sigma )=\sigma _m[j_c(\sigma )/j_d](b/d)$ with
$j_d=d_n\dot{\epsilon} _0$ and $\dot{\epsilon} _0=b\rho v_0$ being
the maximum values of the dislocation current density and the plastic
deformation rate, respectively.
\begin{figure*}[t]
\centerline{\includegraphics[width=9.cm,clip=true]{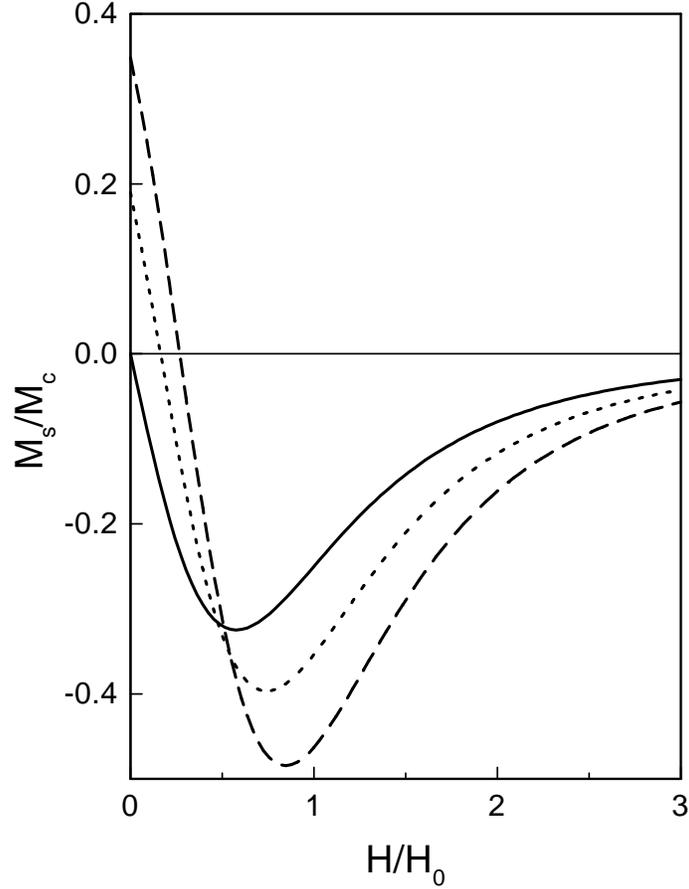}}
\caption{ The reduced magnetization $M_s/M_c$ as a function of the
reduced applied magnetic field $H/H_0$, according to Eq.(25) for
different values of reduced applied stress: $\sigma /\sigma _c=0$
(solid line), $\sigma /\sigma _c=0.01$ (dotted line), and $\sigma
/\sigma _c=0.05$ (dashed line).}
\end{figure*}
According to the recent experiments [43], the tunneling dominated
critical current $I_c$ (and its density $j_c$) in HTCS ceramics was
found to exponentially increase under compressive stress, viz.
$I_c(\sigma )=I_c(0)e^{\beta \sigma }$ with $\beta \simeq 1/\sigma
_m$. Specifically, the critical current at $\sigma =9 kbar$ was found
to be three times higher its value at $\sigma =1.5 kbar$, clearly
indicating a weak-links-mediated origin of the phenomenon (in the
best defect-free thin films this ratio is controlled by the stress
induced modifications of the carrier number density and practically
never exceeds a few percents [42]). Strictly speaking, the critical
current will also change (decrease) with applied magnetic field.
However, for the fields under discussion (see below) in the first
approximation this effect can be neglected. In view of Eq.(25),
dependence of $I_c$ on $\sigma$ will lead to a rather strong
piezomagnetic effects. Indeed, Fig.3 shows changes of the initial
stress-free diamagnetic magnetization $M_s/M_c$ (solid line) under an
applied stress $\sigma /\sigma _c$. Here $M_c\equiv M_0(0)$ and
$\sigma _c\equiv \sigma _0(0)$ (see below for estimates). As we see,
already relatively small values of an applied stress render a low
field Meissner phase strongly paramagnetic (dotted and dashed lines)
simultaneously increasing the maximum of the magnetization and
shifting it towards higher magnetic fields. According to Eq.(25), the
initially diamagnetic Meissner effect turns paramagnetic as soon as
the piezomagnetic contribution $H^{*}(\sigma )$ exceeds an applied
magnetic field $H$. To see whether this can actually happen in a real
material, let us estimate a magnitude of the piezomagnetic field
$H^{*}$. Typically, for HTCS ceramics $S\approx 10\mu m^2$, leading
to $H_0\simeq 1G$. To estimate the needed value of the dislocation
current density $j_d$, we turn to the available experimental data.
According to Ref.39, a rather strong polarization under compressive
pressure $\sigma /\sigma _m \simeq 0.1$ was observed in $YBCO$
ceramic samples at $T=77K$ yielding $d_n=10^2C/m^2$ for the
piezoelectric coefficient. Usually, for GBJJs $\dot{\epsilon}
_0\simeq 10^{-2}s^{-1}$, and $b\simeq 10nm$ leading to
$j_d=d_n\dot{\epsilon} _0\simeq 1A/m^2$ for the maximum dislocation
current density. Using the typical values of the critical current
density $j_c(\sigma )=10^4A/m^2$ (found [43] for $\sigma /\sigma
_m\simeq 0.1$) and grain size $d\simeq 1\mu m$, we arrive at the
following estimate of the piezomagnetic field $H^{*}\simeq
10^{-2}H_0$. Thus, the predicted stress induced paramagnetic Meissner
effect should be observable for applied magnetic fields $H\simeq
10^{-2}H_0\simeq 0.01G$ which correspond to the region where the
original PME was first registered [1-3]. In turn, the piezoelectric
coefficient $d_n$ is related to a charge $Q$ in the GBJJ as [44]
$d_n=(Q/S)(d/b)^2$. Given the above-obtained estimates, we get
$Q\simeq 10^{-13}C$ for an effective charge accumulated by the GBs.
Notice that the above values of the aplied stress $\sigma $ and the
resulting effective charge $Q$ correspond (via the so-called
electroplastic effect [44]) to an equivalent applied electric field
$E=b^2\sigma /Q\simeq 10^7V/m$ at which rather pronounced
electric-field induced effects in HTCS have been recently observed
[17-21].

Besides, according to Ref.43 the Josephson projected area $S$
slightly decreases under pressure thus leading to some increase of
the characteristic field $H_0=\Phi _0/S$. In view of Eq.(25), it
means that a smaller compression stress is needed to actually reverse
the sign of the induced magnetization $M_s$. Furthermore, if an
unloaded granular superconductor already exhibits the PME, due to the
orbital currents induced spontaneous magnetization resulting from an
initial phase difference $\phi ^0_{i,j}=2\pi r$ in Eq.(23) with
fractional $r$ (in particular, $r=1/2$ corresponds to the so-called
[2,3,21] $\pi$-type state), then according to our scenario this
effect will either be further enhanced by applying a compression
(with $\sigma
>0$) or will disappear under a strong enough extension (with $\sigma
<0$). Given the markedly different mechanisms and scales of stress induced
changes in defect-free thin films [42] and weak-links-ridden ceramics
[43], it should be possible to experimentally register the suggested
here piezophase effects.

\vspace{8mm} \leftline{\bf ACKNOWLEDGMENTS} \vspace{5mm}

This work was conceived and partially done during my stay at the
Universidade Federal de S\~ao Carlos (Brazil) where it was funded by
the Brazilian Agency FAPESP (Projeto 2000/04187-8). I thank Wilson
Ortiz and Fernando Araujo-Moreira for hospitality and stimulating
discussions on the subject. I am also indebted to Anant Narlikar for
his invitation to make this contribution for the current volume of
the Studies.

\vspace{16mm}

\vspace{30mm}




\end{document}